\documentclass[showpacs,amsmath,amssymb,twocolumn,pra]{revtex4-1}

\usepackage[dvips]{graphicx} 
\usepackage{amsfonts}
\usepackage{amssymb}
\usepackage{amscd}
\usepackage{amsmath}    
\usepackage{enumerate}
\usepackage{epsfig}
\usepackage{subfigure}
\usepackage{bm}
\usepackage{xcolor}
\usepackage{amsthm}
\usepackage{framed}
\usepackage{multirow}
\usepackage{mathrsfs,amssymb}
\usepackage{latexsym} 
\usepackage{framed}
\usepackage{physics}
\usepackage{epstopdf}
\usepackage[skins]{tcolorbox}
\newtcolorbox[auto counter]{tbox}[2][]{%
    enhanced, float=hbt, drop fuzzy shadow southeast,
    colback=white!5!white, colframe=white!50!black,
    width= .97\columnwidth,sharp corners, boxrule=0.8pt,
    title={Table \thetcbcounter: #2}, #1
}

\begin{document}
\title{Numerical Framework for Semi-Device-Independent Quantum Random Number Generators}

\begin{abstract}
Quantum random number generator (QRNG) is one of the most widely applied branches in quantum cryptography. Among all QRNG schemes, semi-device-independent (semi-DI) QRNG is quite promising, achieving high randomness generation rate with few assumptions on the devices. For the central task of a QRNG study -- security analysis, numerical approaches become popular for its generality to various semi-DI QRNG schemes.  
Here we formulate a numerical framework for the finite-size security of general semi-DI QRNGs, which gives a secure lower bound of the finite-size randomness generation rate against general attacks. We consider a simple example of an optical semi-DI QRNG as an application of our framework.  
\end{abstract}

\author{Hongyi Zhou}
\affiliation{State Key Lab of Processors, Institute of Computing Technology, Chinese Academy of Sciences, 100190, Beijing, China.}

\maketitle
\section{Introduction}
Current pseudo-random numbers suffer from predictions by machine learnings \cite{truong2018machine}. Quantum randomness, in contrast, is in principle unpredictable, which is guaranteed by the fundamental law of quantum mechanics. The machine that can output true randomness is called a quantum random number generator \cite{MaQRNGReview16,RMPQRNGReview16}, which consists of an entropy source and a measurement device that extracts the randomness. Substantial efforts have been devoted to designing various QRNG for higher randomness generation rate and security level, namely, fewer assumptions necessary to guarantee the security of the output.
Among all current QRNG designs, the trusted-device QRNG can achieve the highest randomness generation rate up to 68 Gbps \cite{Nie15}. While the design with the highest security level is called device-independent (DI) QRNG where neither the source nor the measurement device is trusted. Its randomness generation rate is limited to hundreds of bits per second \cite{liu2018device,liu2021device,shalm2021device,PhysRevLett.126.050503}.

In general, a higher security level means a sacrifice on the randomness generation rate. A feasible compromise is called semi-device-independent (semi-DI) QRNGs \cite{cao2015loss,Brunner15,Banik15,Ma16,Nie16,vsupic2017measurement,bischof2017measurement,PhysRevA.100.062338,avesani2018source,li2019QRNG,smith2019simple,rusca2020fast,PhysRevLett.118.060503,PhysRevApplied.15.034034}. In a semi-DI QRNG, some partial knowledge of the devices is permitted, which can dramatically improve the randomness generation rate compared with a fully DI-QRNG. Two main categories of semi-DI QRNGs are source-independent \cite{Ma16,avesani2018source,smith2019simple,PhysRevX.10.041048} and measurement-device-independent QRNGs \cite{cao2015loss,Nie16,bischof2017measurement}, where we leave only half of the device, source or measurement device untrusted, respectively. We will focus on these two categories throughout this paper.


The central task of a QRNG scheme is to prove the security, i.e., randomness quantification in the presence of the adversary Eve, who tries to extract as much information as possible in the output random numbers. Most previous semi-QRNG works analyze the security analytically, where specific techniques such as uncertainty relation \cite{Ma16,PhysRevLett.118.060503}, tomography \cite{cao2015loss}, and dimension witness \cite{Brunner15} are applied. Recently, the numerical method was proposed, which transforms the randomness quantification problem into semidefinite programming (SDP) problems. Compared with analytical methods, numerical ones are quite general to various QRNG schemes, without the requirement of specific techniques. Moreover, tighter randomness lower bounds can usually be obtained by numerical methods when suitable constraints are chosen.

Most current numerical works focus on the asymptotic security, which provides an asymptotic randomness generation rate as an approximation of the performance in an infinite experiment time. For the finite-size security, identically independent distributed (i.i.d) outcomes are often assumed for the convenience of analyzing the statistical fluctuations. Such assumptions are not valid for practical implementations of a QRNG since the outcomes can be correlated under the most general attacks. Therefore, the i.i.d assumption need to be removed to achieve the ultimate goal of the security analysis --- proving the finite-size security against general attacks. 

In this work, we propose a numerical framework for semi-device-independent QRNGs, i.e., source-independent and measurement-device-independent QRNGs, which completes the finite-size security analysis against general attacks. Our framework is based on semidefinite programming (SDP) problems. Different from previous numerical works that care about the optimal value of the primal SDP problem, we consider the solution to the dual problem, which are some real-valued Lagrange multipliers enabling us to construct an operator inequality. Combining proper concentration inequalities, we can calculate the lower bound of randomness in finite-size case. As an example, we consider a simple optical setting which can be viewed as a source-independent or a measurement-device-independent QRNG by leaving the rest part untrusted. By applying our framework, we make the finite-size security analysis for both cases. It turns out that the randomness generation rate outperforms previous works with similar settings.  

\section{Numerical Framework for Source-Independent QRNGs}
\subsection{Protocol}
\begin{figure}[hbt]
\centering
\resizebox{8cm}{!}{\includegraphics{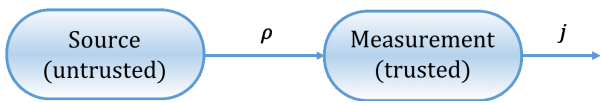}}
\caption{Illustration of a SI-QRNG.} 
\label{fig:sisetting}
\end{figure}

The protocol of a general SI-QRNG is described as follows and illustrated in Fig.~\ref{fig:sisetting}. 
\begin{enumerate}
	\item The untrusted source sends an unknown quantum state to the measurement device possessed by the user Alice.
	\item The measurement device characterized by a set of positive-operator-valued measure (POVM) $\{M_j\}_{j=1}^n$ outputs a measurement result $j$. With probability $p_{\mathrm{sig}}$, Alice determines the round is a signal round. Otherwise it is a test round.
	\item After repeating steps 1-2 for $N_{\mathrm{tot}}$ rounds, Alice records the number of outcomes for each $j$. The number of the outcomes $j$ in the test rounds is denoted as $N_j$.
	\item Alice calculates the upper bound of the number of successful guesses $N_{\mathrm{guess}}^U(N_{\mathrm{tot}},\{N_j\}_j,\epsilon)$ according to Eq.~\eqref{eq:guessuSI}, where $\epsilon$ is the failure probability of estimating the number of successful guesses. Then she performs the post-processing to extract a final random number length of
	\begin{equation}\label{eq:lengthSI}
	N_{\mathrm{fin}} = -\left(N_{\mathrm{tot}}-\sum_j N_j\right)\log_2\left(\frac{N_{\mathrm{guess}}^U(N_{\mathrm{tot}},\{N_j\}_j,\epsilon)}{N_{\mathrm{tot}}-\sum_j N_j}\right)
	\end{equation}
\end{enumerate}

\subsection{Randomness Quantification in Asymptotic Limit}
A general quantum state $\rho$ can be expressed as a pure-state decomposition,
\begin{equation}\label{eq:statedecomp}
\rho = \sum_i p_i \ket{\psi_i}\bra{\psi_i}.
\end{equation}
The way of decomposition in Eq.~\eqref{eq:statedecomp} is not unique, which is determined by the adversary Eve. 
For a given decomposition, Eve's guessing strategy reduces to a classical one, i.e., the guessing probability equals to the maximum probability among all the outcomes. Then considering all possible decompositions, the guessing probability is a maximization,
\begin{equation}\label{eq:sidecomp}
p_{\mathrm{guess}} = \max_{p_i, \ket{\psi_i}} \sum_i p_i \max_j \mathrm{tr}(\ket{\psi_i}\bra{\psi_i}M_j).
\end{equation} 
which is usually difficult to calculate since the number of terms is not fixed. Here we assume the dimensions of the quantum state $\rho$ and POVM element are the same, denoted by $d$. Though the dimension of the source in a SI-QRNG is unknown, its components out of the $d$-dimensional space can be regarded as possessed by Eve. Then according to the Carathéodory's theorem, for a $d$-dimensional density matrix, Eq.~\eqref{eq:guessingprob} can be expressed as a convex combination of at most $d^2$ terms.

To simplify the calculation, we consider a re-grouping into $n$-groups $(n\leq d^2)$ where $n$ is the number of POVM elements. The pure-state component $\ket{\psi_i}$ in the $k$-th group $S_{k}$ satisfies $\max_j \mathrm{tr}(\ket{\psi_i}\bra{\psi_i}M_j) =\mathrm{tr}(\ket{\psi_i}\bra{\psi_i} M_{k})$. After the re-grouping, Eq.~\eqref{eq:statedecomp} can be rewritten as
\begin{equation}
p_{\mathrm{guess}} = \max_{\rho_k} \sum_{k=1}^n \mathrm{tr}(\rho_k M_k),
\end{equation}
where $\rho_k$ is a sub-normalized quantum state $\rho_k =\sum_{i\in S_k}p_i\ket{\psi_i}\bra{\psi_i}$. Then the randomness quantification can be formulated as the following SDP problem,
\begin{equation}\label{eq:SDPprimalSI}
\begin{aligned}
& \max_{\rho_k} \sum_{k=1}^n \mathrm{tr}\left(\rho_k M_k\right) \\
\mathrm{s.t.} \quad & \mathrm{tr}\left(M_j\sum_{j=1}^n\rho_j\right) = \nu_j \\
& \mathrm{tr}\left(\sum_{k=1}^n \rho_k\right) =1 \\
& \rho_k \succeq 0, 
\end{aligned}
\end{equation}
where the first constraint means the unknown source should be compatible with the experimental statistics $\nu_j:= \lim_{N_{\mathrm{tot}}\rightarrow \infty} N_j/(N_{\mathrm{tot}}(1-p_{\mathrm{sig}}))$, the second constraint is the normalization condition. Here we do not use the condition $\max_j \mathrm{tr}(\rho_kM_j) =\mathrm{tr}(\rho_k M_{k})$ since it is automatically satisfied, which can be proved by contradiction. Suppose $\rho_1^*$, $\rho_2^*$, $\cdots$, $\rho_n^*$ are the solutions of Eq.~\eqref{eq:SDPprimalSI} and the corresponding optimal value of the primal problem Eq.~\eqref{eq:SDPprimalSI} is $p^*$. If $\mathrm{tr}(\rho_1^* M_1)\leq \mathrm{tr}(\rho_1^* M_2)$, we can always find another set of solution $\rho_1^{**} =0$, $\rho_2^{**}=\rho_1^* +\rho_2^*$, $\rho_k^{**}=\rho_k^*$ $(k\in\{3,4,\cdots,n\})$. Then the corresponding optimal value $p^{**} \geq p^*$, which leads to a contradiction with the assumption that $p^*$ is the optimal value.

By calculating the SDP problem in Eq.~\eqref{eq:SDPprimalSI}, one can obtain the upper bound of the guessing probability $p_{\mathrm{guess}}^U$ in asymptotic limit. The randomness is quantified by the conditional min-entropy,
\begin{equation}\label{eq:randomnessasym}
H_{\mathrm{min}}(A|E) = -\log_2 p_{\mathrm{guess}}.
\end{equation}

\subsection{Finite-size Analysis}
The asymptotic security is a special case of the finite-size security, i.e., when $N_{\mathrm{tot}}\rightarrow \infty$ the finite-size security reduces to an asymptotic one. We consider the dual problem of Eq.~\eqref{eq:SDPprimalSI},
\begin{equation}\label{eq:SDPdualSI}
\begin{aligned}
& \min_{\vec{\lambda}} -\sum_{j=1}^n \lambda_j \nu_j -\lambda_{n+1} \\
\mathrm{s.t.} \quad & M_k +\sum_{j=1}^n \lambda_j M_j +\lambda_{n+1} I \preceq 0, 
\end{aligned}
\end{equation}
where $\vec{\lambda} = (\lambda_1,\lambda_2, \cdots \lambda_{n+1})$ is the dual variable. The dual problem enables us to obtain some operator inequalities. First we make an estimation of $\nu_j$ by assuming a physical model, which are called nominal values denoted by $\nu_j^{\mathrm{nom}}$. We substitute $\nu_j^{\mathrm{nom}}$ into the dual problem in Eq.~\eqref{eq:SDPdualSI} and obtain the solution $\vec{\lambda}^* = (\lambda^*_1,\lambda^*_2, \cdots \lambda^*_{n+1})$. Then we obtain $n$ operator inequalities, 
\begin{equation}\label{eq:opineq}
M_k +\sum_{j=1}^n \lambda^*_j M_j +\lambda^*_{n+1} I \preceq 0, 
\end{equation}
which are independent of the input states, i.e., we can choose arbitrary quantum states $\rho$ and $\mathrm{tr}[\rho(M_k +\sum_{j=1}^n \lambda^*_j M_j +\lambda^*_{n+1} I)]\leq0$ always holds. A key difference between our framework and previous numerical works is that we use $\nu_j^{\mathrm{nom}}$ instead of $\nu_j$ in the process above. We intuitively explain the reason. To make sure the finite-size randomness generation rate is independent of Eve's attack strategy, the bounded difference in the concentration inequalities should be independent of the actual experimental outcomes (as shown in Eq.~\eqref{eq:boundeddiff}). Applying $\nu_j^{\mathrm{nom}}$ will lead to a looser upper bound of the guessing probability, which will not affect the security. One can refer to \cite{zhou2021numerical} for detailed discussions.

In the $u$-th round, we introduce random variables $\chi^{(u)}$ whose values are taken following the rule in Table.~\ref{tab:randomvariablesi}. We consider a post-measurement quantum state $\rho^{F_{u-1}} = \sum_k \rho_k^{F_{u-1}}$ satisfying that the measurement results of the first $u-1$ rounds coincides with the values from $\chi^{(1)}$ to $\chi^{(u-1)}$. Recalling Eq.~\eqref{eq:opineq}, we have 
\begin{equation}\label{eq:expineq}
\begin{aligned}
&\sum_{k=1}^n \mathrm{tr}\left[\rho_k^{F_{u-1}}\left(M_k +\sum_{j=1}^n \lambda^*_j M_j +\lambda^*_{n+1} I\right)\right] \\
= & E(\chi^{(u)}|F_{u-1}) + \lambda^*_{n+1}   \\
\leq & 0.
\end{aligned}
\end{equation}

\begin{table*}[hbt]
\begin{tabular}{c|c}
\hline
Value of $\chi^{(u)}$ & Event of the $u$-th round \\
\hline
$\frac{1}{p_{\mathrm{sig}}}$  & generation round, Eve successfully guesses the output \\
$\frac{\lambda^*_{j}}{(1-p_{\mathrm{sig}})}$  & test round, measurement device outputs $j$\\
$0$  & other cases \\
\hline
\end{tabular}
\caption{The value of random variable $\chi^{(u)}$ in a SI-QNRG.}
\label{tab:randomvariablesi}
\end{table*}

By applying a concentration inequality to $\chi^{(u)}$, we have an inequality in the following form
\begin{equation}\label{eq:azumaineq}
\sum_{u=1}^{N_{\mathrm{tot}}} \chi^{(u)} \leq \sum_{u=1}^{N_{\mathrm{tot}}} E(\chi^{(u)}|F_{u-1}) + \Delta(N_{\mathrm{tot}},\epsilon),
\end{equation}
which holds with probability at least $1-\epsilon$. Specifically, if we apply Azuma's inequality, the explicit form of $\Delta$ is give by
\begin{equation}
\begin{aligned}
\Delta(N_{\mathrm{tot}},\epsilon) = \sqrt{-2N_{\mathrm{tot}}c^2 \ln\epsilon},
\end{aligned}
\end{equation}
where $c$ is the bounded difference of the martingale $\sigma_t = \sum_{u=1}^t \chi^{(u)} -\sum_{u=1}^t E(\chi^{(u)}|F_{u-1})$,
\begin{equation}\label{eq:boundeddiff}
c  = 2\max\left(\frac{1}{p_{\mathrm{sig}}}, \frac{\lambda_1^*}{1-p_{\mathrm{sig}}}, \frac{\lambda_2^*}{1-p_{\mathrm{sig}}}, \cdots, \frac{\lambda_n^*}{1-p_{\mathrm{sig}}}\right)
\end{equation}
Combining Eqs.~\eqref{eq:expineq} and \eqref{eq:azumaineq}, we can obtain the upper bound of the number of successful guesses $N_{\mathrm{guess}}^U(N_{\mathrm{tot}},\{N_j\}_j,\epsilon)$ given by
\begin{equation}\label{eq:guessuSI}
\begin{aligned}
&N_{\mathrm{guess}}^U(N_{\mathrm{tot}},\{N_j\}_j,\epsilon) :=  \\
&p_{\mathrm{sig}}N_{\mathrm{tot}}\left(-\sum_{j=1}^n \frac{\lambda_j^* N_j}{N_{\mathrm{tot}}(1-p_{\mathrm{sig}})}-\lambda_{n+1}^*  + \Delta(N_{\mathrm{tot}},\epsilon) \right)
\end{aligned}
\end{equation}
which holds with probability at least $1-\epsilon$.

\section{Numerical Framework for Measurement-Device-Independent QRNGs}
\subsection{Protocol}

\begin{figure}[hbt]
\centering
\resizebox{8cm}{!}{\includegraphics{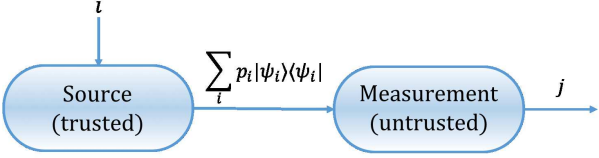}}
\caption{Illustration of a MDI-QRNG.} 
\label{fig:mdisetting}
\end{figure}

The protocol of a general MDI-QRNG is described as follows and illustrated in Fig.~\ref{fig:mdisetting}. 
\begin{enumerate}
	\item The user Alice randomly prepares a pure state $\ket{\psi_i}$ with probability $p_i$ $(i\in\{1,2,\cdots,m\})$ to an unknown measurement device. With probability $p_{\mathrm{sig}}$, Alice determines the round is a signal round. Otherwise it is a test round.
	\item The measurement device outputs a measurement result $j$ $(j\in\{1,2,\cdots,n\})$. 
	\item After repeating steps 1-2 for $N_{\mathrm{tot}}$ rounds, Alice records the number of outcomes for each $j$ given the test state $\ket{\psi_i}$ in the test rounds, denoted by $N_{j|i}$.
	\item Alice calculates the upper bound of the number of successful guesses $N_{\mathrm{guess}}^U(N_{\mathrm{tot}},\{N_{j|i}\}_{i,j},\epsilon)$ according to Eq.~\eqref{eq:guessuMDI}, where $\epsilon$ is the failure probability of estimating the number of successful guesses. Then she performs the post-processing to extract a final random number length of
	\begin{equation}\label{eq:lengthMDI}
	\begin{aligned}
	N_{\mathrm{fin}}& = -\left(N_{\mathrm{tot}}-\sum_{i=1}^m \sum_{j=1}^n N_{j|i}\right) \\
	& \times \log_2\left(\frac{N_{\mathrm{guess}}^U(N_{\mathrm{tot}},\{N_{j|i}\}_{i,j},\epsilon)}{N_{\mathrm{tot}}-\sum_{i=1}^m \sum_{j=1}^n N_{j|i}}\right).
	\end{aligned}
	\end{equation}
\end{enumerate}

Here are some remarks on the protocol. The dimension of the unknown POVM is set to be the same as $\rho = \sum_{i=1}^m p_i \ket{\psi_i}\bra{\psi_i}$. According to Naimark's dilation theorem, the actual measurement can be a projective measurement on larger system including Alice's signal on ${\cal H}_A$ and Eve's ancillary system on ${\cal H}_E$. Eve's measurement on ${\cal H}_E$ will determine the decomposition of the unknown POVM, which will be explained in detail later.

\subsection{Randomness Quantification in Asymptotic Limit}
To quantify the output randomness of a POVM, we consider a decomposition into extremal POVMs \cite{d2005classical}. For simplicity, we consider a single test state $\ket{\psi}$. According to \cite{dai2022intrinsic}, the randomness is given by a convex-roof form,
\begin{equation}\label{eq:POVMrandomness}
\begin{aligned}
R = \min_{q^\eta,\mathbf{M}^\eta} \sum_{\eta} q^\eta R(\ket{\psi}\bra{\psi}, \mathbf{M}^\eta),
\end{aligned}
\end{equation}
where $\sum_\eta q^\eta =1$, $\mathbf{M}^\eta$ is a set of extremal POVM, and $R(\ket{\psi}\bra{\psi}, \mathbf{M}^\eta)$ is given by
\begin{equation}\label{eq:extPOVMrandomness}
\begin{aligned}
R(\ket{\psi}\bra{\psi}, \mathbf{M}^\eta) & = R(\ket{\psi}\bra{\psi}\otimes \ket{0}\bra{0}_E, \mathbf{P}^\eta)\\
& = -\log_2 \max_j \mathrm{tr}\left[\left(\ket{\psi}\bra{\psi}\otimes \ket{0}\bra{0}_E\right)P_j^\eta \right] \\
& = -\log_2\max_j \mathrm{tr}\left(\ket{\psi}\bra{\psi} M_j^\eta\right).
\end{aligned}
\end{equation}
Now we explain Eq.~\eqref{eq:extPOVMrandomness}. The first equality is based on the property of a set of extremal POVM that all the Naimark dilation gives the same randomness (Thm. 1 in \cite{dai2022intrinsic}). Then we consider a canonical dilation with randomness $R(\ket{\psi}\bra{\psi}\otimes \ket{0}\bra{0}_E, \mathbf{P}^\eta)$, where $\mathbf{P}^\eta=\{P_j^\eta\}_j$ is a projective measurement. Then the randomness is given by the classical min-entropy in the rhs of the second equality. The third equality comes from the property of a Naimark dilation. Suppose the optimal decomposition that minimizes Eq.~\eqref{eq:extPOVMrandomness} is $(q^{\eta*}, \mathbf{M}^{\eta*})$, then we have the following relation according to the concavity of a logarithmic function
\begin{equation}
R_1 = -\log_2 \sum_\eta q^{\eta*} \max_j  \mathrm{tr}\left(\ket{\psi}\bra{\psi} M_j^{\eta*}\right) \leq R.
\end{equation}
We define the guessing probability as
\begin{equation}\label{eq:guessingprob}
p_{\mathrm{guess}} = \max_{q^\eta,\mathbf{M}^\eta} \sum_\eta q^\eta \max_j  \mathrm{tr}\left(\ket{\psi}\bra{\psi} M_j^{\eta}\right).
\end{equation}
Then
\begin{equation}
-\log_2 p_{\mathrm{guess}} \leq R_1 \leq R,
\end{equation}
which means we can obtain a valid lower bound of the randomness by calculating the guessing probability. In general, the number of terms in Eq.~\eqref{eq:guessingprob} is not fixed. Again by using Carathéodory's theorem, for a $d$-dimensional Hilbert space that the POVM acts on, Eq.~\eqref{eq:guessingprob} can be expressed as a convex combination of at most $nd^2$ terms. Then we let the number of terms in the summation be $nd^2$ without loss of generality. We further consider a re-grouping of these terms, i.e., we can always divide these POVM components into $n$ groups. The terms in the $l$-th group $S_l$ satisfy $\max_j \mathrm{tr}(\ket{\psi}\bra{\psi}M_j^\eta) = \mathrm{tr}(\ket{\psi}\bra{\psi}M_l^\eta)$, then Eq.~\eqref{eq:guessingprob} is rewritten as 
\begin{equation}
\begin{aligned}\label{eq:probguess_combined}
p_{\mathrm{guess}} & = \max_{q^\eta,\mathbf{M}^\eta} \sum_l \sum_{q^\eta \in S_l} q^\eta \mathrm{tr}\left(\ket{\psi}\bra{\psi} M_l^{\eta}\right) \\
& = \max_{\Lambda^l_j} \sum_l \mathrm{tr}\left(\ket{\psi}\bra{\psi}\Lambda^l_l\right),
\end{aligned}
\end{equation}
where the probabilities $q^\eta$ is absorbed into the POVM elements. Then the randomness quantification can be formulated as a SDP problem
\begin{equation}\label{eq:SDPprimalMDI_singleteststate}
\begin{aligned}
&\max_{\Lambda^l_j} \sum_l \mathrm{tr}\left(\ket{\psi}\bra{\psi}\Lambda^l_l\right)\\
\mathrm{s.t.} \quad  &\sum_{l=1}^n  \mathrm{tr}\left(\ket{\psi}\bra{\psi} \Lambda_j^l\right)=\nu_j\\
&\sum_{j=1}^n\sum_{l=1}^n  \Lambda_j^l=I \\
&\Lambda_{j}^l\succeq 0,
\end{aligned}
\end{equation} 
where $\nu_j$ has the same definition as that in Eq.~\eqref{eq:SDPprimalSI}. This can be easily extended into the cases where $m$ $(m\leq 2 \log_nd +1)$ test states are prepared and the POVM is divided into $n^m$ groups. Then Eq.~\eqref{eq:SDPprimalMDI_singleteststate} is generalized into
\begin{equation}\label{eq:SDPprimalMDI}
\begin{aligned}
&\max_{\Lambda^{l_1l_2\dots l_m}_j} \sum_{i=1}^m p_i \sum_{l_1=1}^n \sum_{l_2=1}^n \cdots \sum_{l_m=1}^n \mathrm{tr}\left(\ket{\psi_i}\bra{\psi_i}\Lambda_{l_i}^{l_1l_2\dots l_m}\right)\\
\mathrm{s.t.} \quad  &\sum_{l_1=1}^n \sum_{l_2=1}^n \cdots \sum_{l_m=1}^n \mathrm{tr}\left(\ket{\psi_i}\bra{\psi_i} \Lambda_j^{l_1l_2\dots l_m}\right)=\nu_{j|i} \\
&\sum_{j=1}^n\sum_{l_1=1}^n \sum_{l_2=1}^n \cdots \sum_{l_m=1}^n \Lambda_j^{l_1l_2\dots l_m}=I \\
&\Lambda_{j}^{l_1l_2\dots l_m} \succeq 0.
\end{aligned}
\end{equation}
For both Eqs.~\eqref{eq:SDPprimalMDI_singleteststate} and~\eqref{eq:SDPprimalMDI}, the first constraint means the unknown POVM should be compatible with the experimental statistics $\nu_{j|i}:=\lim_{N_{\mathrm{tot}}\rightarrow \infty} N_{j|i}/(N_{\mathrm{tot}}(1-p_{\mathrm{sig}}))$, the second constraint is the normalization condition and the last one is the positive semi-definite condition.
Here we remark that the framework for MDI-QRNGs has a similar form as that in \cite{brask2017megahertz} and thus is
also applicable to semi-device-independent QRNGs with overlap bounds on the sources, where one can construct density matrix of the source states parametrized by the overlap bounds \cite{brask2017megahertz}.

\subsection{Finite-size Analysis}
For the finite-size analysis, we consider the dual problem of Eq.~\eqref{eq:SDPprimalMDI}, 
\begin{equation}\label{eq:SDPdualMDI}
\begin{aligned}
&\min_{H^{l_1l_2\dots l_m},\eta_{ij}} - \sum_{i=1}^m \sum_{j=1}^n \eta_{ij} \nu_{j|i} \\
\mathrm{s.t.} \quad &  H^{l_1l_2\dots l_m} = (H^{l_1l_2\dots l_m})^\dag  \\
& \sum_{i=1}^m  \ket{\psi_i}\bra{\psi_i} \left(p_i\sum_{k=1}^n \delta_{\lambda_i, j} \delta_{k,j}+\eta_{ij}\right) \\
&+H^{l_1l_2\dots l_m}-\mathrm{tr}(H^{l_1l_2\dots l_m})I \preceq 0.
\end{aligned}
\end{equation}
We also calculate the nominal values $\nu_{j|i}^{\mathrm{nom}}$ before experiments. Suppose the solution to the dual problem Eq.~\eqref{eq:SDPdualMDI} is $\eta^*_{ij}$ and $H^{l_1l_2\dots l_m,*}$. We have the following operator inequalities, 
\begin{equation}
\begin{aligned}
&\sum_{i=1}^m  \ket{\psi_i}\bra{\psi_i} \left(p_i\sum_{k=1}^n \delta_{\lambda_i, j} \delta_{k,j}+\eta^*_{ij}\right) \\
&+H^{l_1l_2\dots l_m,*}-\mathrm{tr}\left(H^{l_1l_2\dots l_m,*}\right)I \preceq 0,
\end{aligned}
\end{equation}
which holds for arbitrary $\Lambda_{j}^{l_1l_2\dots l_m}$. 

We also introduce a random variable $\chi^{(u)}$ satisfying the rule in Table~\ref{tab:randomvariable}.
\begin{table*}[hbt]
\begin{tabular}{c|c}
\hline
Value of $\chi^{(u)}$ & Event of the $u$-th round \\
\hline
$\frac{1}{p_{\mathrm{sig}}}$  & generation round, Eve's successful guessing \\
$\frac{\eta^*_{ij}}{(1-p_{\mathrm{sig}})p_i}$  & test round, $\ket{\psi_i}$ is prepared and measurement outcome is $j$\\
$0$  & other cases. \\
\hline
\end{tabular}
\caption{The value of random variable $\chi^{(u)}$ in a MDI-QNRG.}
\label{tab:randomvariable}
\end{table*}
Then there always exists a set of $\Lambda_{j}^{l_1l_2\dots l_m, F_{u-1}}$ such that
\begin{equation}\label{eq:operatordc}
\begin{aligned}
 & \sum_{l_1 =1}^n\sum_{l_2 =1}^n \cdots \sum_{l_m=1}^n \sum_{i=1}^m p_i \mathrm{tr}\left(\ket{\psi_i}\bra{\psi_i} \Lambda_{l_i}^{l_1l_2\dots l_m, F_{u-1}}\right) \\
 &+ \sum_{j=1}^n \sum_{i=1}^m \eta^*_{ij}\mathrm{tr} \left(\sum_{l_1 =1}^n\sum_{l_2 =1}^n \cdots \sum_{l_m=1}^n \ket{\psi_i}\bra{\psi_i}\Lambda_{j}^{l_1l_2\dots l_m, F_{u-1}}\right) \\
= & E(\chi^{(u)}|F_{u-1}) \\
\leq & 0,
\end{aligned}
\end{equation}

By applying concentration inequalities which hold with probability at least $1-\epsilon$ to $\chi^{(u)}$, 
we can calculate the upper bound of successful guesses following the same procedure of the SI-QRNG case
\begin{equation}\label{eq:guessuMDI}
\begin{aligned}
&N_{\mathrm{guess}}^U(N_{\mathrm{tot}},\{N_{j|i}\}_{i,j},\epsilon) := \\& p_{\mathrm{sig}}N_{\mathrm{tot}} \left( \sum_{i=1}^m \sum_{j=1}^n \frac{\eta^*_{ij}}{N_{\mathrm{tot}}(1-p_{\mathrm{sig}})p_i}N_{j|i} + \Delta(N_{\mathrm{tot}},\epsilon) \right),
\end{aligned}
\end{equation}
which holds with probability at least $1-\epsilon$.




\section{Example}
\begin{figure}[hbt]
\centering
\resizebox{8cm}{!}{\includegraphics{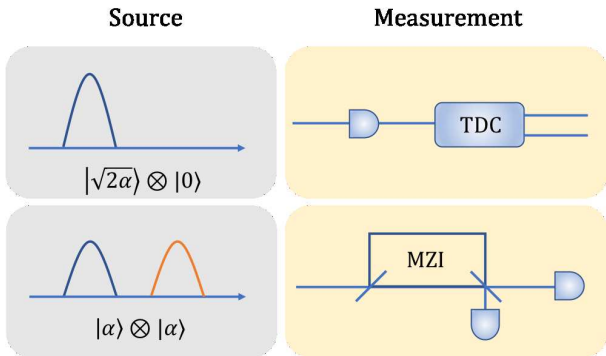}}
\caption{Settings of the source and measurement in a time-bin phase-encoding optical system. MZI: Mach-Zehnder Interferometer; TDC: time-to-digital converter.} 
\label{fig:timebin}
\end{figure}

Now we give an example of how to apply our framework. We consider a time-bin phase-encoding optical system composed of a weak coherent state source and several threshold detectors. A coherent state $\ket{\alpha}$ is given by a superposition of Fock states $\ket{n}$,
\begin{equation}
\ket{\alpha} = e^{-\frac{|\alpha|^2}{2}}\sum_{n=0}^\infty \frac{\alpha^n}{\sqrt{n!}}\ket{n}.
\end{equation}
The source randomly sends two coherent states $\ket{\sqrt{2}\alpha}_1 \otimes \ket{0}_2$ and $\ket{\alpha}_1\otimes \ket{\alpha}_2$ with probabilities $p_s$ and $1-p_s$, respectively. The subscript $1$ and $2$ represent the label of the time bins. 
The measurement devices randomly switches between a $Z$-basis measurement $\{\ket{0}\bra{0},\ket{1}\bra{1}\}$ and a $X$-basis measurement $\{\ket{+}\bra{+},\ket{-}\bra{-}\}$ with probabilities $p_z$ and $1-p_z$, respectively. In the time-bin phase-encoding optical system, the $Z$-basis measurement is realized by a threshold detector followed by a time-to-digital converter (TDC) while the $X$-basis measurement is realized by an interferometer followed by two threshold detectors. We illustrate the experiment settings in Fig.~\ref{fig:timebin}.

\begin{figure*}[hbt]
\centering
   \begin{minipage}[t]{0.48\textwidth}
   \centering
 \includegraphics[width=8cm]{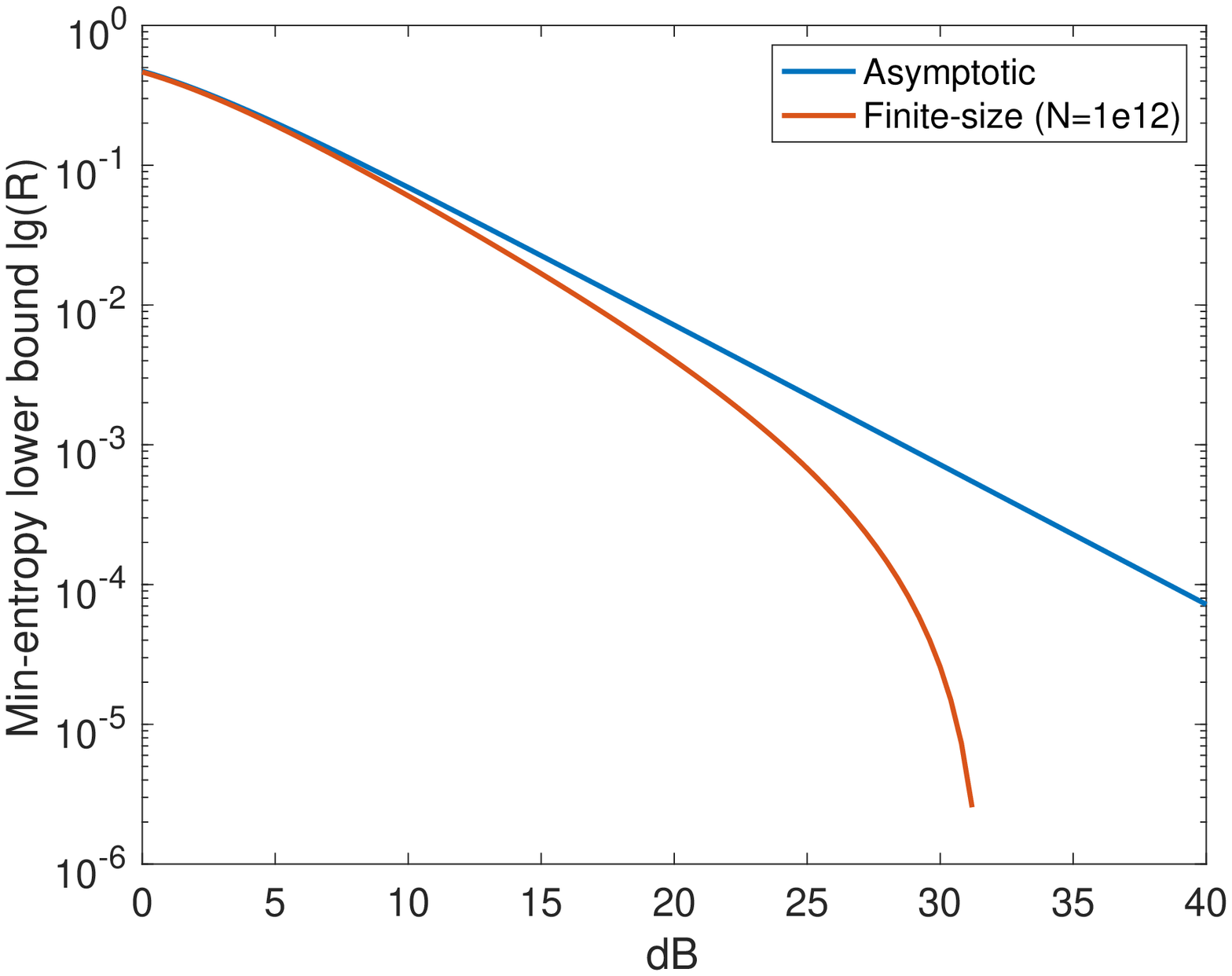}
   \end{minipage}
   \begin{minipage}[t]{0.48\textwidth}
   \centering
  \includegraphics[width=8cm]{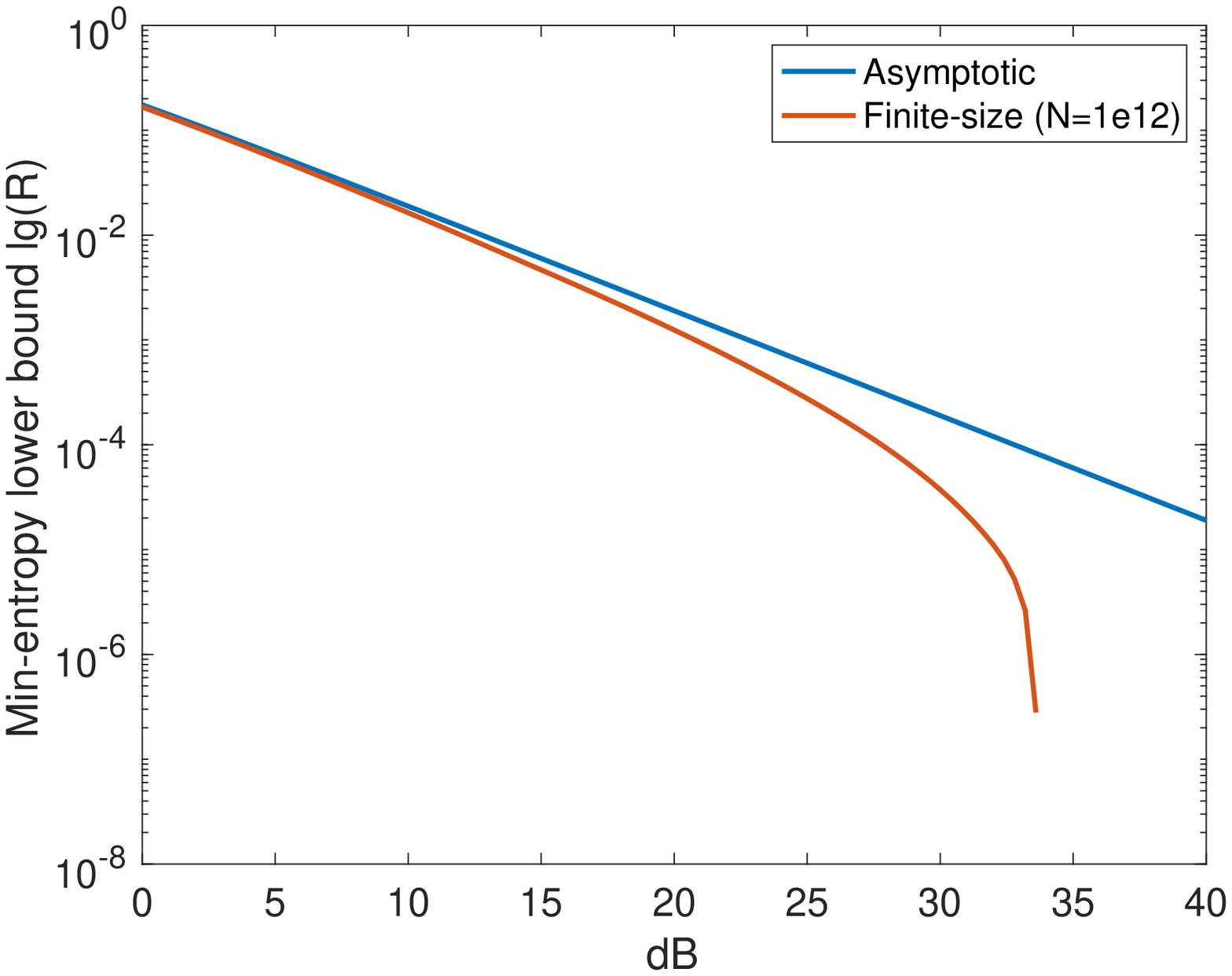}
   \end{minipage}

\caption{Simulations of the finite-size min-entropy lower bound versus loss when the setting is regarded as a SI-QRNG (left) and a MDI-QRNG (right). The intensities are optimized.} 
\label{fig:simulation}
\end{figure*}

The channel is assumed to be a lossy channel characterized by the transmittance $\eta$. It transforms a coherent state $\ket{\alpha}$ into $\ket{\sqrt{\eta}\alpha}$. 
The parameters are set as follows.
The number of total rounds is $N_{\mathrm{tot}}=10^{12}$; the total failure probability of applying concentration inequalities is $\epsilon = 10^{-10}$; the dark count rate of the detectors is $p_d =10^{-8}$; the probability of choosing $\rho_1$ is $p_s = 0.5$; the probability of choosing $Z$ basis measurement is $p_z =0.5$.
The probability of choosing signal rounds as $p_{\mathrm{sig}}$ needs to be optimized. This setting can be either viewed as a SI-QRNG or a MDI-QRNG. The simulation results for both cases are shown in Fig.~\ref{fig:simulation}. The min-entropy lower bound calculated by our framework can be improved by orders of magnitudes compared with previous works with similar settings \cite{cao2015loss,Ma16}.

\subsection{Randomness Quantification As a SI-QRNG}
In a SI-QRNG, we need the characterization of the measurement settings and leave the source uncharacterized. Thanks to the squashing model \cite{PhysRevLett.101.093601}, the measurement setting can be described as a three-dimensional POVM.
\begin{equation}
\begin{aligned}
M_1 &= p_z\ket{0}\bra{0}\oplus 0 \\
M_2 & = p_z\ket{1}\bra{1}\oplus 0 \\
M_3 & =(1-p_z) \ket{+}\bra{+} \oplus 0 \\
M_4 & =(1-p_z) \ket{+}\bra{+} \oplus 0 \\
M_5 & = I - \sum_{i=1}^4 M_i.
\end{aligned}
\end{equation}
Then we simulate $q_j^{\mathrm{nom}}$ $(j\in\{1,2,3,4\})$. The formulas are given in Appendix~\ref{app:simulation}.
We can calculate $\vec{\lambda}^*$ by substituting $q_j^{\mathrm{nom}}$ into Eq.~\eqref{eq:SDPdualSI}. Assuming $N_j = N_{\mathrm{tot}}(1-p_{\mathrm{sig}})q_j^{\mathrm{nom}}$, we can calculate the upper bound of successful guessing by Eq.~\eqref{eq:guessuSI} and the final random number length by Eq.~\eqref{eq:lengthSI}.

\subsection{Randomness Quantification As a MDI-QRNG}
In a MDI-QRNG, one cannot distinguish a $Z$ and $X$ basis measurement. Then the measurement device outputs $1$ for $\ket{0}\bra{0}$ and $\ket{+}\bra{+}$, outputs $2$ for $\ket{1}\bra{1}$ and $\ket{-}\bra{-}$, and outputs $3$ for inconclusive results, which forms a coarse-grained POVM $\{M_{j}\}$ $(j \in\{1,2,3\})$. The source is rewritten in a set of canonical basis $\{\ket{\phi},\ket{\phi^\perp}\}$,
\begin{equation}
\begin{aligned}
\ket{\sqrt{2}\alpha}_1 \otimes \ket{0}_2 & = \ket{\phi} \\
\ket{\alpha}_1\otimes \ket{\alpha}_2& = e^{-(2+\sqrt{2})|\alpha|^2}\ket{\phi}\\&\;\;+\sqrt{1-e^{-2(2-\sqrt{2})|\alpha|^2}}\ket{\phi^\perp}.
\end{aligned}
\end{equation}

We simulate $q_{j|i}^{\mathrm{nom}}$ $(i\in\{1,2\},j\in\{1,2,3\})$ with formulas given in Appendix~\ref{app:simulation}.
 Then we can calculate $\eta_{ij}^*$ by substituting $q_{j|i}^{\mathrm{nom}}$ into Eq.~\eqref{eq:SDPdualMDI}. Assuming $N_{j|i} = N_{\mathrm{tot}}p_i(1-p_{\mathrm{sig}})q_{j|i}^{\mathrm{nom}}$, we can calculate the upper bound of successful guessing by Eq.~\eqref{eq:guessuMDI} and the final random number length by Eq.~\eqref{eq:lengthMDI}. 

\section{conclusion}

In conclusion, we formulate a numerical framework for semi-DI QRNGs which gives finite-size security against general attacks. Our example shows that the numerical method can give tighter lower bounds of randomness generation rate. For future directions, we plan to explore the following problems. First, it is still possible to further relax assumptions in the current framework. An example is the MDI-QRNG with overlap bound on the source \cite{brask2017megahertz}. Then it is interesting to generalize the framework such that it can deal with the security analysis of a larger class of QRNG schemes, for example, MDI-QRNG with partial information on the source and SI-QRNG with partial information on the POVM. Second, if we make the maximum relaxation on the assumptions, the semi-DI QRNG becomes a DI one. Then we would like to explore whether the semi-DI QRNG and DI-QRNG can be further unified into a single framework. Finally, the numerical approach is naturally suitable for unstructured protocol designs, which enables us to deal with various device imperfections in practical implementations.

\section{Acknowledgments}
I thank M. Koashi, Y. Nie, T. Sasaki and X. Zhang for enlightening discussions. This work was supported in part by the National Natural Science Foundation of China Grants No. 61832003, 61872334, 61801459, and the Strategic Priority Research Program of Chinese Academy of Sciences Grant No. XDB28000000.

\onecolumngrid

\appendix
\section{Simulation formulas}\label{app:simulation}
We list the simulation formulas as follows. Suppose the intensity of the source is $\mu$, i.e., $\mu=2|\alpha|^2$. For a SI-QRNG, we consider the probabilities of outputting $j$ $(j\in \{1,2,3,4,5\})$ given the $i$-th state $(i\in\{1,2\})$.
\begin{equation}
\begin{aligned}
p(1|\rho_1) & =  p_z \left((1-(1-p_d)e^{-\mu\eta})(1-p_d)+0.5p_d(1-(1-p_d)e^{-\mu\eta})\right)\\
p(2|\rho_1) & = p_z \left( p_d(1-p_d)e^{-\mu\eta} +0.5p_d(1-(1-p_d)e^{-\mu\eta}) \right) \\
p(3|\rho_1) & =  (1-p_z) \left( (1-(1-pd)e^{-\mu\eta/2})(1-p_d)e^{-\mu\eta/2}+0.5(1-(1-p_d)e^{-\mu\eta/2})^2\right)\\
p(4|\rho_1) & =  (1-p_z) \left( (1-(1-pd)e^{-\mu\eta/2})(1-p_d)e^{-\mu\eta/2}+0.5(1-(1-p_d)e^{-\mu\eta/2})^2\right) \\
p(5|\rho_1) & =  p_z\left((1-p_d)^2 e^{-\mu\eta})+(1-p_z)((1-p_d)^2e^{-\mu\eta}\right) \\
p(1|\rho_2) & =  p_z((1-(1-p_d)e^{-\mu\eta/2})(1-p_d)e^{-\mu\eta/2}+0.5(1-(1-p_d)e^{-\mu\eta/2})^2) \\
p(2|\rho_2) & =  p_z((1-(1-p_d)e^{-\mu\eta/2})(1-p_d)e^{-\mu\eta/2}+0.5(1-(1-p_d)e^{-\mu\eta/2})^2) \\
p(3|\rho_2) & =  (1-p_z)(p_d(1-p_d)e^{-\mu\eta}+0.5p_d(1-(1-pd)e^{-\mu\eta})) \\
p(4|\rho_2) & =  (1-p_z)((1-(1-p_d)e^{-\mu\eta})(1-pd)+0.5p_d(1-(1-p_d)e^{-\mu\eta})) \\
p(5|\rho_2) & =  p_z((1-p_d)^2e^{-\mu\eta})+(1-p_z)((1-p_d)^2e^{-\mu\eta}).\\
\end{aligned}
\end{equation}

Then $q_j^{\mathrm{nom}}$ is given by
\begin{equation}\label{eq:simulationsi}
q_j^{\mathrm{nom}} = p_s p(j|\rho_1) + (1-p_s) p(j|\rho_2).
\end{equation}

For a MDI-QRNG, the conditional probabilities $q_{j|i}^{\mathrm{nom}}$ is given by
\begin{equation}\label{eq:simulationmdi}
\begin{aligned}
 q_{1|1}^{\mathrm{nom}} & = p_z(1-(1-p_d)e^{-\mu\eta})(1-p_d)+(1-p_z)\left((1-(1-p_d)e^{-\mu\eta/2})(1-p_d)e^{-\mu\eta/2}\right) \\
 q_{2|1}^{\mathrm{nom}} & = p_z p_d(1-p_d)e^{-\mu\eta} +(1-p_z)\left((1-(1-p_d)e^{-\mu\eta/2})(1-p_d)e^{-\mu\eta/2}\right) \\
 q_{3|1}^{\mathrm{nom}} & = 1 - q_{1|1}^{\mathrm{nom}} - q_{2|1}^{\mathrm{nom}} \\
 q_{1|2}^{\mathrm{nom}} & = p_z p_d(1-p_d)e^{-\mu\eta} +(1-p_z)\left((1-(1-p_d)e^{-\mu\eta/2})(1-p_d)e^{-\mu\eta/2}\right) \\
 q_{2|2}^{\mathrm{nom}} & = p_z(1-(1-p_d)e^{-\mu\eta})(1-p_d)+(1-p_z)\left((1-(1-p_d)e^{-\mu\eta/2})(1-p_d)e^{-\mu\eta/2}\right) \\
 q_{3|2}^{\mathrm{nom}} & = 1 - q_{1|2}^{\mathrm{nom}} - q_{2|2}^{\mathrm{nom}}. \\
\end{aligned}
\end{equation}



\twocolumngrid
\bibliography{bibsemidiqrng}
\end{document}